\documentclass[a4paper]{jpconf}
\usepackage{amsmath,commath,siunitx}
\usepackage{graphicx,xcolor}
\usepackage{indentfirst}

\usepackage[square,sort&compress,numbers]{natbib}
\newbox\ASYbox
\newdimen\ASYdimen
\long\def\ASYbase#1#2{\leavevmode\setbox\ASYbox=\hbox{#1}\setbox\ASYbox=\hbox{#2}\lower\ASYdimen\box\ASYbox}
\long\def\ASYaligned(#1,#2)(#3,#4)#5#6#7{\leavevmode\setbox\ASYbox=\hbox{#7}\setbox\ASYbox\hbox{\ASYdimen=\ht\ASYbox\advance\ASYdimen by\dp\ASYbox\kern#3\wd\ASYbox\raise#4\ASYdimen\box\ASYbox}\setbox\ASYbox=\hbox{#5\wd\ASYbox 0pt\dp\ASYbox 0pt\ht\ASYbox 0pt\box\ASYbox#6}\hbox to 0pt{\kern#1pt\raise#2pt\box\ASYbox\hss}}\long\def\ASYalignT(#1,#2)(#3,#4)#5#6{\ASYaligned(#1,#2)(#3,#4){}{}{#6}}
\long\def\ASYalign(#1,#2)(#3,#4)#5{\ASYaligned(#1,#2)(#3,#4){}{}{#5}}
\def\ASYraw#1{#1}

\begin{document}
\title{Heavy flavor electron $R_\text{AA}$ and $v_2$ in event-by-event relativistic hydrodynamics}
\author{Caio A~G~Prado\textsuperscript{1}, Mauro R~Cosentino\textsuperscript{1,2}, Marcelo G~Munhoz\textsuperscript{1}, Jorge Noronha\textsuperscript{1,3} and Alexandre A~P~Suaide\textsuperscript{1}}
\address{\textsuperscript{1} Instituto de F\'{i}sica, Universidade de S\~{a}o Paulo, C.P. 66318, 05315-970 S\~{a}o Paulo, SP, Brazil}
\address{\textsuperscript{2} Centro de Ci\^{e}ncias Naturais e Humanas, Universidade Federal do ABC, Av. dos Estados 5001, Bairro Santa Terezina, 09210-508 Santo Andr\'{e}, SP, Brazil}
\address{\textsuperscript{3} Department of Physics, Columbia University, 538 West 120th Street, New York, NY 10027, USA}
\ead{caio.prado@usp.br mr.cosentino@gmail.com munhoz@if.usp.br noronha@if.usp.br suaide@if.usp.br}

\begin{abstract}
  In this work we investigate how event-by-event hydrodynamics
  fluctuations affect the nuclear suppression factor and elliptic flow
  of heavy flavor mesons and non-photonic electrons. We use a 2D+1
  Lagrangian ideal hydrodynamic code~\cite{Andrade2013,Andrade2014a} on
  an event-by-event basis in order to compute local temperature and flow
  profiles. Using a strong coupling inspired energy loss
  parametrization~\cite{Gubser2006} on top of the evolving space-time energy
  density distributions we are
  able to propagate the heavy quarks inside the medium until the
  freeze-out temperature is reached and a \textsc{Pythia}~\cite{SJOSTRAND2008} modeling of
  hadronization takes place. The resulting D$^0$ and heavy-flavor
  electron yield is compared with recent experimental data for
  $R_\text{AA}$ and $v_2$ from the \textsc{Star} and \textsc{Phenix}
  collaborations~\cite{Adare2007,Abelev2007,Adamczyk}. In addition we present preditions
  for the higher order Fourier harmonic coefficients $v_3(p_T)$
  of heavy-flavor electrons at \textsc{Rhic}'s
  $\sqrt{S_\text{NN}} = \SI{200}{GeV}$ collisions.
\end{abstract}

\section{Introduction}

Central Au+Au collisions at \textsc{Rhic} and Pb+Pb at \textsc{Lhc} exhibit a strong particle
suppression when compared to p+p collisions as well as anisotropic flow. The
suppression is usually related with jet quenching or energy loss of partons
inside the quark-gluon-plasma (\textsc{qgp}) whereas the flow might be due to
lumps of higher density inside the medium due to initial fluctuations. Furthermore while the \textsc{qgp}
dynamics may affect differently the expansion of these high-density spots in
comparison with the rest of the plasma it can affect higher harmonic orders
of anisotropic flow such as $v_3$.

We aim to study these effects of the medium on the heavy quarks suppression
during their evolution inside the \textsc{qgp}. In order to achieve this goal
we performed computational simulations of the evolution of heavy quarks. The
simulation consists of a sampled quark drawn from an initial condition energy
density profile. The quark is then evolved inside the medium while the medium
itself expands. The initial $p_T$ distribution is given by First-Order
Next-to-Leading Logs (\textsc{fonll}~\cite{Cacciari2005}) calculation. After
evolution quarks reach the freeze-out temperature and hadronize. The
resulting meson decays into electrons we can analyse and compare with
experimental data of electron $R_\text{AA}$ and $v_2$.

\section{Simulation}

This simulation takes for granted the factorization on \textsc{qcd} in order
to follow a modular paradigm. This allows one to replace part of the
calculation without affecting the other ones, for instance, one could change
the energy loss model in order to study a different one. We can summarize the
simulation with the following modules:
\begin{itemize}
  \item Initial Conditions;
  \item Hydrodynamics;
  \item Energy loss model;
  \item Fragmentation;
  \item Meson decay.
\end{itemize}

Also, the simulation only takes into account the effects of the medium over
the quark probes and not the other way around.

\subsection{Initial Conditions}

Initial conditions are constructed using a modified version of the
\textsc{Phobos} Glauber Monte Carlo~\cite{Loizides2014} code with default
parameters, and selecting the appropriate centrality range. From the nucleons participants in the collision given by
the \textsc{Phobos} simulation we can create an energy density profile for
central rapidity by attributing gaussian distributions for each wounded
nucleon and binary collision~\cite{Nonaka2007} with width given in terms of
the inelastic nucleon-nucleon cross section~\cite{Loizides2014}. The normalization factor
which arises from this method is calculated by matching the maximum initial
average temperature with literature
values~\cite{Luzum2009,Luzum2010,Roy2011b,Bhatt2011}.

\subsection{Hydrodynamics}

The evolution of the medium is performed by an implementation of the Smoothed
Particles Hydrodynamics (\textsc{sph})~\cite{Aguiar2000,Andrade2013}
algorithm. We use a longitudinal expansion with boost
invariance~\cite{Hwa1974,Bjorken1983} and set the initial transverse velocity
to zero while also considering it to be independent of the rapidity. In
addition we assume the baryon chemical potential to be zero. The equation of
state \textsc{eos} \textsc{s}95n-v1~\cite{Huovinen2010} is used in this
work~\cite{Andrade2014a}.

The evolution starts from $\tau = \SI{1.0}{fm\per c}$ and we set the
freeze-out temperature to \SI{140}{MeV}~\cite{Luzum2009,Luzum2010}. The
smoothing \textsc{sph} parameter is set to $h = \SI{0.3}{fm}$ in order to
perform the simulation in a doable time while still preserving the important
structures of the initial conditions. The evolution goes until the complete
decoupling of the particles in the medium following the Cooper-Frye
prescription~\cite{Cooper1974}. The hydrodynamics is performed separately
from the quark evolution.

\subsection{Energy Loss Model}

We draw quarks from the medium using the initial energy density profile as
probability distribution for the initial position. From a \textsc{fonll}
calculation~\cite{Cacciari1998,Cacciari2001,Cacciari2005} we set the initial
momentum of the quark with uniformly distributed $\varphi$ direction.

The quarks are propagated by a numerical integration of the energy loss over
step distance. The integration is performed locally in the medium frame so
Lorentz transformations are performed every step and this accounts for some
quarks gaining energy while being pushed by the medium.

The parametrization used in these simulations for the energy loss is given
by:
\begin{equation}\label{eq:egyloss}
  \od{E}{x} \propto v \gamma(v) T^2,
\end{equation}
where $v$ is the quark velocity, $\gamma(v)$ is the Lorentz factor and $T$ is
the medium temperature.
A similar expression is obtained by using \textsc{a}d\textsc{s}/\textsc{cft}
correspondence and a classical test string approximation to calculate the
drag force on an external quark moving in a thermal plasma~\cite{Gubser2006}
and a Langevin simulation to describe the diffusion dynamics of heavy quarks
in \textsc{qgp}~\cite{Akamatsu2009}.

The scale factor of the parametrization is fitted against experimental data,
we use D$^0$ meson $R_\text{AA}$ for the charm quarks factor and the electron
$R_\text{AA}$ for the bottom one.

\subsection{Fragmentation, decay and output}

After the propagation is performed and the quarks have reached the freeze-out
temperature we use Peterson fragmentation function in order to hadronize the
quarks which then decay into electrons using the semi-leptonic channels through
\textsc{Pythia}8~\cite{SJOSTRAND2008}.

The final output is the electron distribution after all quarks have
hadronized and decayed. By analysing these electron spectra we can obtain
heavy flavor electron $R_\text{AA}$ and $v_2$ in order to compare the results
with experimental data.

\section{Results}

\begin{figure}[hbt]
  \begin{minipage}[b]{0.48\textwidth}
    \centering
                \setlength{\unitlength}{1pt}
\makeatletter\let\ASYencoding\f@encoding\let\ASYfamily\f@family\let\ASYseries\f@series\let\ASYshape\f@shape\makeatother{\catcode`"=12\includegraphics{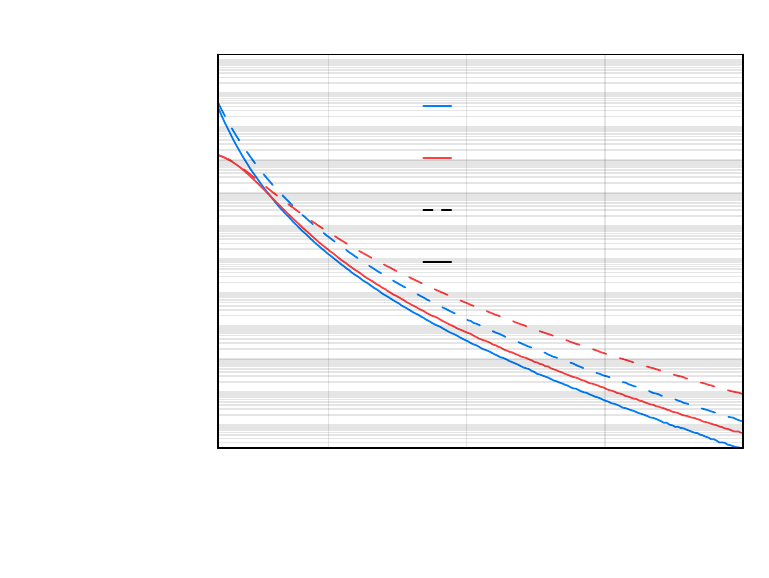}}\begin{picture}( 222.132818, 167.035840)\end{picture}\kern -222.132818pt\definecolor{ASYcolor}{gray}{0.000000}\color{ASYcolor}
\fontsize{12.045000}{14.454000}\selectfont
\usefont{\ASYencoding}{\ASYfamily}{\ASYseries}{\ASYshape}\ASYalign(-127.181642,28.135020)(-0.500000,-1.000000){\vphantom{$10^4$}$5$}\definecolor{ASYcolor}{gray}{0.000000}\color{ASYcolor}
\fontsize{12.045000}{14.454000}\selectfont
\ASYalign(-87.247950,28.135020)(-0.500000,-1.000000){\vphantom{$10^4$}$10$}\definecolor{ASYcolor}{gray}{0.000000}\color{ASYcolor}
\fontsize{12.045000}{14.454000}\selectfont
\ASYalign(-47.314257,28.135020)(-0.500000,-1.000000){\vphantom{$10^4$}$15$}\definecolor{ASYcolor}{gray}{0.000000}\color{ASYcolor}
\fontsize{12.045000}{14.454000}\selectfont
\ASYalign(-7.380565,28.135020)(-0.500000,-1.000000){\vphantom{$10^4$}$20$}\definecolor{ASYcolor}{gray}{0.000000}\color{ASYcolor}
\fontsize{12.045000}{14.454000}\selectfont
\ASYalign(-83.254581,13.505625)(-0.500000,-0.750000){$p_T$ (\si{GeV})}\definecolor{ASYcolor}{gray}{0.000000}\color{ASYcolor}
\fontsize{12.045000}{14.454000}\selectfont
\ASYalign(-168.432648,44.115904)(-1.000000,-0.500000){\vphantom{$10^4$}$10^{-4}$}\definecolor{ASYcolor}{gray}{0.000000}\color{ASYcolor}
\fontsize{12.045000}{14.454000}\selectfont
\ASYalign(-168.432648,63.307560)(-1.000000,-0.500000){\vphantom{$10^4$}$10^{-2}$}\definecolor{ASYcolor}{gray}{0.000000}\color{ASYcolor}
\fontsize{12.045000}{14.454000}\selectfont
\ASYalign(-168.432648,82.499215)(-1.000000,-0.500000){\vphantom{$10^4$}$10^{0}$}\definecolor{ASYcolor}{gray}{0.000000}\color{ASYcolor}
\fontsize{12.045000}{14.454000}\selectfont
\ASYalign(-168.432648,101.690870)(-1.000000,-0.500000){\vphantom{$10^4$}$10^{2}$}\definecolor{ASYcolor}{gray}{0.000000}\color{ASYcolor}
\fontsize{12.045000}{14.454000}\selectfont
\ASYalign(-168.432648,120.882525)(-1.000000,-0.500000){\vphantom{$10^4$}$10^{4}$}\definecolor{ASYcolor}{gray}{0.000000}\color{ASYcolor}
\fontsize{12.045000}{14.454000}\selectfont
\ASYalign(-168.432648,140.074180)(-1.000000,-0.500000){\vphantom{$10^4$}$10^{6}$}\definecolor{ASYcolor}{gray}{0.000000}\color{ASYcolor}
\fontsize{12.045000}{14.454000}\selectfont
\ASYalignT(-211.116923,94.344583)(-0.500000,0.000000){0.000000 1.000000 -1.000000 0.000000}{$\frac{1}{\pi} \left.\frac{\dif \sigma}{\dif p_T^2 \dif y}\right|_{y=0}$ (\si{pb\per GeV^2})}\definecolor{ASYcolor}{rgb}{0.000000,0.500000,1.000000}\color{ASYcolor}
\fontsize{12.045000}{14.454000}\selectfont
\ASYalign(-88.104873,136.336181)(0.000000,-0.500000){Charm}\definecolor{ASYcolor}{rgb}{1.000000,0.250000,0.250000}\color{ASYcolor}
\fontsize{12.045000}{14.454000}\selectfont
\ASYalign(-88.104873,121.324931)(0.000000,-0.500000){Bottom}\definecolor{ASYcolor}{gray}{0.000000}\color{ASYcolor}
\fontsize{12.045000}{14.454000}\selectfont
\ASYalign(-88.104873,106.313681)(0.000000,-0.250000){w/o egy. loss}\definecolor{ASYcolor}{gray}{0.000000}\color{ASYcolor}
\fontsize{12.045000}{14.454000}\selectfont
\ASYalign(-88.104873,91.302431)(0.000000,-0.281251){with egy. loss}\definecolor{ASYcolor}{gray}{0.000000}\color{ASYcolor}
\fontsize{12.045000}{14.454000}\selectfont
\ASYalign(-83.254581,154.863595)(-0.500000,0.218749){\normalsize \textsc{Rhic} electron spectra}     \caption{Electron spectra for \textsc{Rhic} ($\sqrt{S_\text{NN}} =
    \SI{200}{GeV}$) energy comparing charm and bottom flavors.}
    \label{fig:ptrhic}
  \end{minipage}
  \hfill
  \begin{minipage}[b]{0.48\textwidth}
    \centering
                \setlength{\unitlength}{1pt}
\makeatletter\let\ASYencoding\f@encoding\let\ASYfamily\f@family\let\ASYseries\f@series\let\ASYshape\f@shape\makeatother{\catcode`"=12\includegraphics{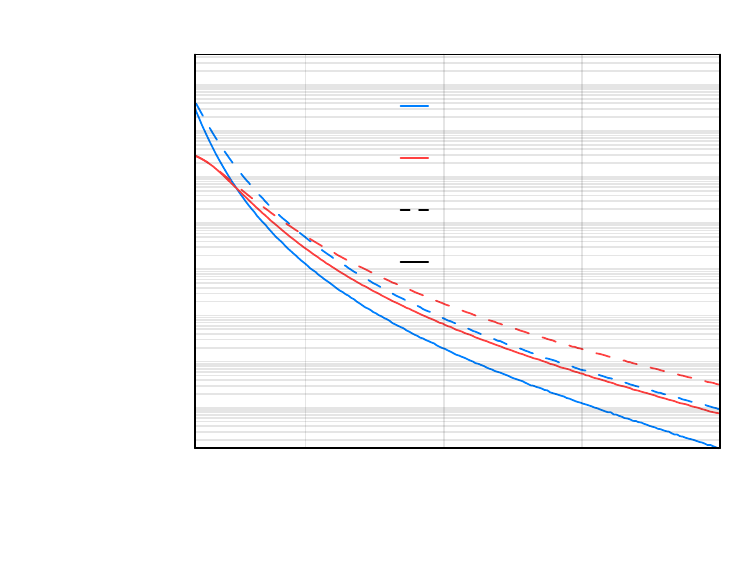}}\begin{picture}( 215.521608, 167.035840)\end{picture}\kern -215.521608pt\definecolor{ASYcolor}{gray}{0.000000}\color{ASYcolor}
\fontsize{12.045000}{14.454000}\selectfont
\usefont{\ASYencoding}{\ASYfamily}{\ASYseries}{\ASYshape}\ASYalign(-127.181642,28.135020)(-0.500000,-1.000000){\vphantom{$10^4$}$5$}\definecolor{ASYcolor}{gray}{0.000000}\color{ASYcolor}
\fontsize{12.045000}{14.454000}\selectfont
\ASYalign(-87.247950,28.135020)(-0.500000,-1.000000){\vphantom{$10^4$}$10$}\definecolor{ASYcolor}{gray}{0.000000}\color{ASYcolor}
\fontsize{12.045000}{14.454000}\selectfont
\ASYalign(-47.314257,28.135020)(-0.500000,-1.000000){\vphantom{$10^4$}$15$}\definecolor{ASYcolor}{gray}{0.000000}\color{ASYcolor}
\fontsize{12.045000}{14.454000}\selectfont
\ASYalign(-7.380565,28.135020)(-0.500000,-1.000000){\vphantom{$10^4$}$20$}\definecolor{ASYcolor}{gray}{0.000000}\color{ASYcolor}
\fontsize{12.045000}{14.454000}\selectfont
\ASYalign(-83.254581,13.505625)(-0.500000,-0.750000){$p_T$ (\si{GeV})}\definecolor{ASYcolor}{gray}{0.000000}\color{ASYcolor}
\fontsize{12.045000}{14.454000}\selectfont
\ASYalign(-168.432648,49.169840)(-1.000000,-0.500000){\vphantom{$10^4$}$10^{0}$}\definecolor{ASYcolor}{gray}{0.000000}\color{ASYcolor}
\fontsize{12.045000}{14.454000}\selectfont
\ASYalign(-168.432648,75.835082)(-1.000000,-0.500000){\vphantom{$10^4$}$10^{2}$}\definecolor{ASYcolor}{gray}{0.000000}\color{ASYcolor}
\fontsize{12.045000}{14.454000}\selectfont
\ASYalign(-168.432648,102.500325)(-1.000000,-0.500000){\vphantom{$10^4$}$10^{4}$}\definecolor{ASYcolor}{gray}{0.000000}\color{ASYcolor}
\fontsize{12.045000}{14.454000}\selectfont
\ASYalign(-168.432648,129.165567)(-1.000000,-0.500000){\vphantom{$10^4$}$10^{6}$}\definecolor{ASYcolor}{gray}{0.000000}\color{ASYcolor}
\fontsize{12.045000}{14.454000}\selectfont
\ASYalignT(-204.505713,94.344583)(-0.500000,0.000000){0.000000 1.000000 -1.000000 0.000000}{$\frac{1}{\pi} \left.\frac{\dif \sigma}{\dif p_T^2 \dif y}\right|_{y=0}$ (\si{pb\per GeV^2})}\definecolor{ASYcolor}{rgb}{0.000000,0.500000,1.000000}\color{ASYcolor}
\fontsize{12.045000}{14.454000}\selectfont
\ASYalign(-88.104873,136.336181)(0.000000,-0.500000){Charm}\definecolor{ASYcolor}{rgb}{1.000000,0.250000,0.250000}\color{ASYcolor}
\fontsize{12.045000}{14.454000}\selectfont
\ASYalign(-88.104873,121.324931)(0.000000,-0.500000){Bottom}\definecolor{ASYcolor}{gray}{0.000000}\color{ASYcolor}
\fontsize{12.045000}{14.454000}\selectfont
\ASYalign(-88.104873,106.313681)(0.000000,-0.250000){w/o egy. loss}\definecolor{ASYcolor}{gray}{0.000000}\color{ASYcolor}
\fontsize{12.045000}{14.454000}\selectfont
\ASYalign(-88.104873,91.302431)(0.000000,-0.281251){with egy. loss}\definecolor{ASYcolor}{gray}{0.000000}\color{ASYcolor}
\fontsize{12.045000}{14.454000}\selectfont
\ASYalign(-83.254581,154.863595)(-0.500000,0.218749){\normalsize \textsc{Lhc} electron spectra}     \caption{Electron spectra for \textsc{Lhc} ($\sqrt{S_\text{NN}} =
    \SI{2.76}{TeV}$) energy comparing charm and bottom flavors.}
    \label{fig:ptlhc}
  \end{minipage}
\end{figure}

We now present some results that were obtained with the simulation.
Figures~\ref{fig:ptrhic} and \ref{fig:ptlhc} show the electron spectra for \textsc{Rhic} and
\textsc{Lhc} energies comparing both bottom and charm flavors with and
without the energy loss. The ratio between them is the nuclear modification
factor $R_\text{AA}$.

\begin{figure}[p]
  \centering
        \setlength{\unitlength}{1pt}
\makeatletter\let\ASYencoding\f@encoding\let\ASYfamily\f@family\let\ASYseries\f@series\let\ASYshape\f@shape\makeatother{\catcode`"=12\includegraphics{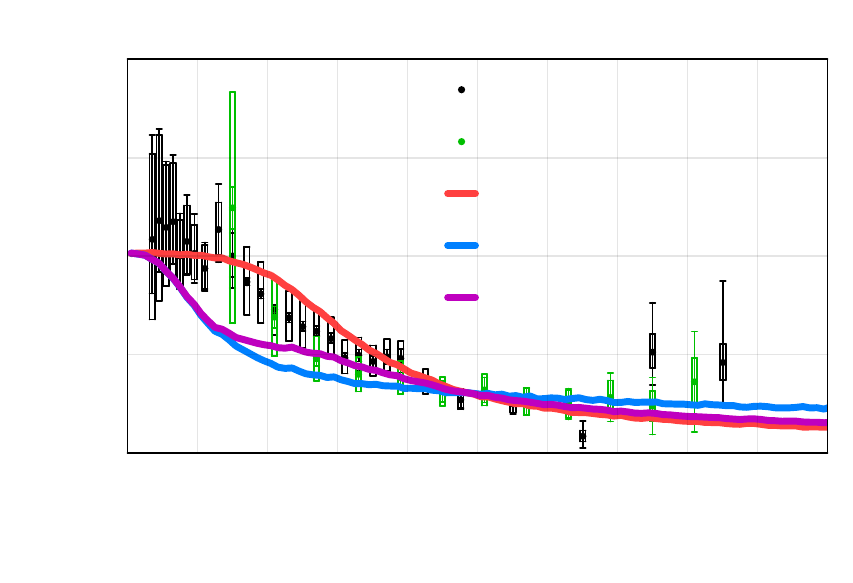}}\begin{picture}( 246.581595, 168.369210)\end{picture}\kern -246.581595pt\definecolor{ASYcolor}{gray}{0.000000}\color{ASYcolor}
\fontsize{12.045000}{14.454000}\selectfont
\usefont{\ASYencoding}{\ASYfamily}{\ASYseries}{\ASYshape}\ASYalign(-209.711274,28.135020)(-0.500000,-1.000000){\vphantom{$10^4$}$0$}\definecolor{ASYcolor}{gray}{0.000000}\color{ASYcolor}
\fontsize{12.045000}{14.454000}\selectfont
\ASYalign(-169.245132,28.135020)(-0.500000,-1.000000){\vphantom{$10^4$}$2$}\definecolor{ASYcolor}{gray}{0.000000}\color{ASYcolor}
\fontsize{12.045000}{14.454000}\selectfont
\ASYalign(-128.778990,28.135020)(-0.500000,-1.000000){\vphantom{$10^4$}$4$}\definecolor{ASYcolor}{gray}{0.000000}\color{ASYcolor}
\fontsize{12.045000}{14.454000}\selectfont
\ASYalign(-88.312848,28.135020)(-0.500000,-1.000000){\vphantom{$10^4$}$6$}\definecolor{ASYcolor}{gray}{0.000000}\color{ASYcolor}
\fontsize{12.045000}{14.454000}\selectfont
\ASYalign(-47.846707,28.135020)(-0.500000,-1.000000){\vphantom{$10^4$}$8$}\definecolor{ASYcolor}{gray}{0.000000}\color{ASYcolor}
\fontsize{12.045000}{14.454000}\selectfont
\ASYalign(-7.380565,28.135020)(-0.500000,-1.000000){\vphantom{$10^4$}$10$}\definecolor{ASYcolor}{gray}{0.000000}\color{ASYcolor}
\fontsize{12.045000}{14.454000}\selectfont
\ASYalign(-108.545919,13.505625)(-0.500000,-0.750000){$p_T$ (\si{GeV})}\definecolor{ASYcolor}{gray}{0.000000}\color{ASYcolor}
\fontsize{12.045000}{14.454000}\selectfont
\ASYalign(-219.015325,37.439071)(-1.000000,-0.500000){\vphantom{$10^4$}$0$}\definecolor{ASYcolor}{gray}{0.000000}\color{ASYcolor}
\fontsize{12.045000}{14.454000}\selectfont
\ASYalign(-219.015325,94.344583)(-1.000000,-0.500000){\vphantom{$10^4$}$1$}\definecolor{ASYcolor}{gray}{0.000000}\color{ASYcolor}
\fontsize{12.045000}{14.454000}\selectfont
\ASYalign(-219.015325,151.250095)(-1.000000,-0.500000){\vphantom{$10^4$}$2$}\definecolor{ASYcolor}{gray}{0.000000}\color{ASYcolor}
\fontsize{12.045000}{14.454000}\selectfont
\ASYalignT(-230.009390,94.344583)(-0.500000,0.194687){0.000000 1.000000 -1.000000 0.000000}{$R_\text{AA}^\text{(hfe)}$}\definecolor{ASYcolor}{gray}{0.000000}\color{ASYcolor}
\fontsize{12.045000}{14.454000}\selectfont
\ASYalign(-105.525238,142.432461)(0.000000,-0.250000){\textsc{Phenix} 0--10\% \cite{Adare2007}}\definecolor{ASYcolor}{gray}{0.000000}\color{ASYcolor}
\fontsize{12.045000}{14.454000}\selectfont
\ASYalign(-105.525238,127.421211)(0.000000,-0.250000){\textsc{Star} 0--5\% \cite{Abelev2007}}\definecolor{ASYcolor}{rgb}{1.000000,0.250000,0.250000}\color{ASYcolor}
\fontsize{12.045000}{14.454000}\selectfont
\ASYalign(-105.525238,112.409961)(0.000000,-0.500000){MC bottom}\definecolor{ASYcolor}{rgb}{0.000000,0.500000,1.000000}\color{ASYcolor}
\fontsize{12.045000}{14.454000}\selectfont
\ASYalign(-105.525238,97.398711)(0.000000,-0.500000){MC charm}\definecolor{ASYcolor}{rgb}{0.750000,0.000000,0.750000}\color{ASYcolor}
\fontsize{12.045000}{14.454000}\selectfont
\ASYalign(-105.525238,82.387461)(0.000000,-0.500000){MC total}\definecolor{ASYcolor}{gray}{0.000000}\color{ASYcolor}
\fontsize{12.045000}{14.454000}\selectfont
\ASYalign(-108.545919,154.863595)(-0.500000,0.179999){\large $R_\text{AA}$ at \textsc{Rhic}}   \caption{Electron nuclear modification factor for each quark flavor and the
    total one in comparison with experiment results for \textsc{Rhic}'s
    energy \SI{200}{GeV}}
  \label{fig:raarhic}
\end{figure}

\begin{figure}[p]
  \centering
        \setlength{\unitlength}{1pt}
\makeatletter\let\ASYencoding\f@encoding\let\ASYfamily\f@family\let\ASYseries\f@series\let\ASYshape\f@shape\makeatother{\catcode`"=12\includegraphics{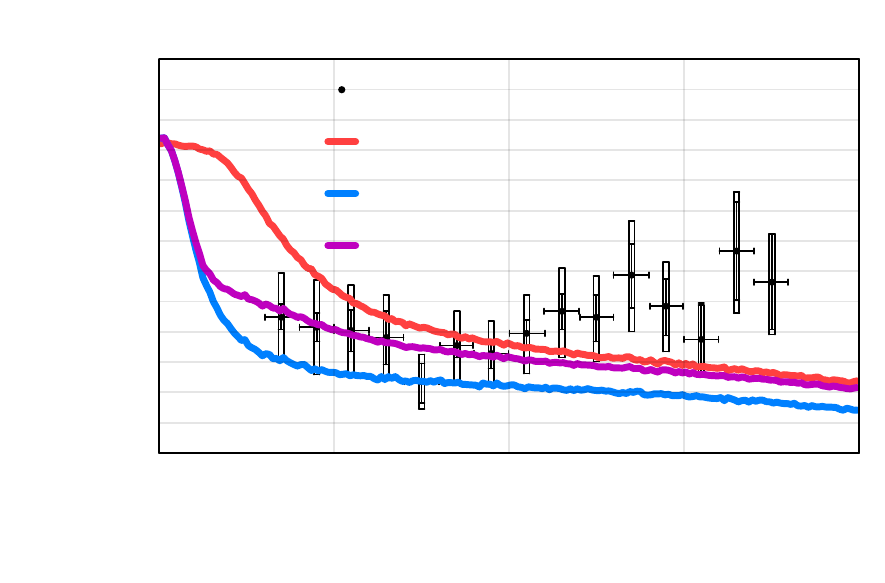}}\begin{picture}( 255.720385, 168.369210)\end{picture}\kern -255.720385pt\definecolor{ASYcolor}{gray}{0.000000}\color{ASYcolor}
\fontsize{12.045000}{14.454000}\selectfont
\usefont{\ASYencoding}{\ASYfamily}{\ASYseries}{\ASYshape}\ASYalignT(-209.711274,28.135020)(-0.500000,-1.000000){1.000000 0.000000 0.000000 1.000000}{\vphantom{$10^4$}$0$}\definecolor{ASYcolor}{gray}{0.000000}\color{ASYcolor}
\fontsize{12.045000}{14.454000}\selectfont
\ASYalignT(-108.545919,28.135020)(-0.500000,-1.000000){1.000000 0.000000 0.000000 1.000000}{\vphantom{$10^4$}$10$}\definecolor{ASYcolor}{gray}{0.000000}\color{ASYcolor}
\fontsize{12.045000}{14.454000}\selectfont
\ASYalignT(-7.380565,28.135020)(-0.500000,-1.000000){1.000000 0.000000 0.000000 1.000000}{\vphantom{$10^4$}$20$}\definecolor{ASYcolor}{gray}{0.000000}\color{ASYcolor}
\fontsize{12.045000}{14.454000}\selectfont
\ASYalignT(-108.545919,13.505625)(-0.500000,-0.750000){1.000000 0.000000 0.000000 1.000000}{$p_T$ (\si{GeV})}\definecolor{ASYcolor}{gray}{0.000000}\color{ASYcolor}
\fontsize{12.045000}{14.454000}\selectfont
\ASYalignT(-219.015325,37.439071)(-1.000000,-0.500000){1.000000 0.000000 0.000000 1.000000}{\vphantom{$10^4$}$0$}\definecolor{ASYcolor}{gray}{0.000000}\color{ASYcolor}
\fontsize{12.045000}{14.454000}\selectfont
\ASYalignT(-219.015325,63.703154)(-1.000000,-0.500000){1.000000 0.000000 0.000000 1.000000}{\vphantom{$10^4$}$0.3$}\definecolor{ASYcolor}{gray}{0.000000}\color{ASYcolor}
\fontsize{12.045000}{14.454000}\selectfont
\ASYalignT(-219.015325,89.967236)(-1.000000,-0.500000){1.000000 0.000000 0.000000 1.000000}{\vphantom{$10^4$}$0.6$}\definecolor{ASYcolor}{gray}{0.000000}\color{ASYcolor}
\fontsize{12.045000}{14.454000}\selectfont
\ASYalignT(-219.015325,116.231318)(-1.000000,-0.500000){1.000000 0.000000 0.000000 1.000000}{\vphantom{$10^4$}$0.9$}\definecolor{ASYcolor}{gray}{0.000000}\color{ASYcolor}
\fontsize{12.045000}{14.454000}\selectfont
\ASYalignT(-219.015325,142.495401)(-1.000000,-0.500000){1.000000 0.000000 0.000000 1.000000}{\vphantom{$10^4$}$1.2$}\definecolor{ASYcolor}{gray}{0.000000}\color{ASYcolor}
\fontsize{12.045000}{14.454000}\selectfont
\ASYalignT(-239.148180,94.344583)(-0.500000,0.194687){0.000000 1.000000 -1.000000 0.000000}{$R_\text{AA}^\text{(hfe)}$}\definecolor{ASYcolor}{gray}{0.000000}\color{ASYcolor}
\fontsize{12.045000}{14.454000}\selectfont
\ASYalign(-149.295288,142.432461)(0.000000,-0.250000){\textsc{Alice} (Prelim.) 0--10\% \cite{Godoy2014}}\definecolor{ASYcolor}{rgb}{1.000000,0.250000,0.250000}\color{ASYcolor}
\fontsize{12.045000}{14.454000}\selectfont
\ASYalign(-149.295288,127.421211)(0.000000,-0.500000){MC bottom}\definecolor{ASYcolor}{rgb}{0.000000,0.500000,1.000000}\color{ASYcolor}
\fontsize{12.045000}{14.454000}\selectfont
\ASYalign(-149.295288,112.409961)(0.000000,-0.500000){MC charm}\definecolor{ASYcolor}{rgb}{0.750000,0.000000,0.750000}\color{ASYcolor}
\fontsize{12.045000}{14.454000}\selectfont
\ASYalign(-149.295288,97.398711)(0.000000,-0.500000){MC total}\definecolor{ASYcolor}{gray}{0.000000}\color{ASYcolor}
\fontsize{12.045000}{14.454000}\selectfont
\ASYalign(-108.545919,154.863595)(-0.500000,0.179999){\large $R_\text{AA}$ at \textsc{Lhc}}   \caption{Electron nuclear modification factor for each quark flavor and the
    total one in comparison with experiment results for \textsc{Lhc}'s
    energy \SI{2.76}{TeV}.}
  \label{fig:raalhc}
\end{figure}

We show the resulting $R_\text{AA}$ calculations for both \textsc{Rhic} and
\textsc{Lhc} energies in Figures~\ref{fig:raarhic} and \ref{fig:raalhc}. The
simulation is performed separately for bottom and charm quark and in order to
obtain the full $R_\text{AA}$ spectra we combine both results weighted by the
\textsc{fonll} cross section calculations. The spectra are compared with experimental data.

\begin{figure}[p]
  \begin{minipage}[b]{0.48\textwidth}
    \centering
                \setlength{\unitlength}{1pt}
\makeatletter\let\ASYencoding\f@encoding\let\ASYfamily\f@family\let\ASYseries\f@series\let\ASYshape\f@shape\makeatother{\catcode`"=12\includegraphics{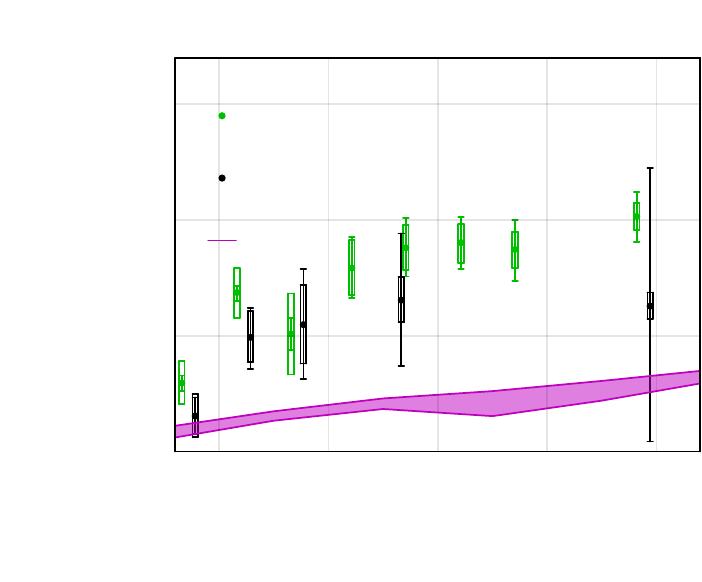}}\begin{picture}( 202.683030, 167.969210)\end{picture}\kern -202.683030pt\definecolor{ASYcolor}{gray}{0.000000}\color{ASYcolor}
\fontsize{12.045000}{14.454000}\selectfont
\usefont{\ASYencoding}{\ASYfamily}{\ASYseries}{\ASYshape}\ASYalignT(-107.739126,28.135020)(-0.500000,-1.000000){1.000000 0.000000 0.000000 1.000000}{\vphantom{$10^4$}$1$}\definecolor{ASYcolor}{gray}{0.000000}\color{ASYcolor}
\fontsize{12.045000}{14.454000}\selectfont
\ASYalignT(-44.510780,28.135020)(-0.500000,-1.000000){1.000000 0.000000 0.000000 1.000000}{\vphantom{$10^4$}$2$}\definecolor{ASYcolor}{gray}{0.000000}\color{ASYcolor}
\fontsize{12.045000}{14.454000}\selectfont
\ASYalignT(-76.124953,13.505625)(-0.500000,-0.750000){1.000000 0.000000 0.000000 1.000000}{$p_T$ (\si{GeV})}\definecolor{ASYcolor}{gray}{0.000000}\color{ASYcolor}
\fontsize{12.045000}{14.454000}\selectfont
\ASYalignT(-161.303020,37.439071)(-1.000000,-0.500000){1.000000 0.000000 0.000000 1.000000}{\vphantom{$10^4$}$0$}\definecolor{ASYcolor}{gray}{0.000000}\color{ASYcolor}
\fontsize{12.045000}{14.454000}\selectfont
\ASYalignT(-161.303020,70.912902)(-1.000000,-0.500000){1.000000 0.000000 0.000000 1.000000}{\vphantom{$10^4$}$0.05$}\definecolor{ASYcolor}{gray}{0.000000}\color{ASYcolor}
\fontsize{12.045000}{14.454000}\selectfont
\ASYalignT(-161.303020,104.386732)(-1.000000,-0.500000){1.000000 0.000000 0.000000 1.000000}{\vphantom{$10^4$}$0.1$}\definecolor{ASYcolor}{gray}{0.000000}\color{ASYcolor}
\fontsize{12.045000}{14.454000}\selectfont
\ASYalignT(-161.303020,137.860563)(-1.000000,-0.500000){1.000000 0.000000 0.000000 1.000000}{\vphantom{$10^4$}$0.15$}\definecolor{ASYcolor}{gray}{0.000000}\color{ASYcolor}
\fontsize{12.045000}{14.454000}\selectfont
\ASYalignT(-187.310815,94.344583)(-0.500000,0.189099){0.000000 1.000000 -1.000000 0.000000}{$v_2^\text{hfe}$}\definecolor{ASYcolor}{gray}{0.000000}\color{ASYcolor}
\fontsize{12.045000}{14.454000}\selectfont
\ASYalign(-130.862898,134.502836)(0.000000,-0.250000){\textsc{Star} 2-corr. \cite{Adamczyk}}\definecolor{ASYcolor}{gray}{0.000000}\color{ASYcolor}
\fontsize{12.045000}{14.454000}\selectfont
\ASYalign(-130.862898,116.489336)(0.000000,-0.250000){\textsc{Star} 4-corr. \cite{Adamczyk}}\definecolor{ASYcolor}{rgb}{0.750000,0.000000,0.750000}\color{ASYcolor}
\fontsize{12.045000}{14.454000}\selectfont
\ASYalign(-130.862898,98.475836)(0.000000,-0.500000){Monte Carlo}\definecolor{ASYcolor}{gray}{0.000000}\color{ASYcolor}
\fontsize{12.045000}{14.454000}\selectfont
\ASYalign(-76.124953,154.863595)(-0.500000,0.068966){\large \textsc{Rhic} $0$--$60$\%}     \caption{Calculation of electron $v_2$ for \textsc{Rhic} energy.}
    \label{fig:v2rhic}
  \end{minipage}
  \hfill
  \begin{minipage}[b]{0.48\textwidth}
    \centering
                \setlength{\unitlength}{1pt}
\makeatletter\let\ASYencoding\f@encoding\let\ASYfamily\f@family\let\ASYseries\f@series\let\ASYshape\f@shape\makeatother{\catcode`"=12\includegraphics{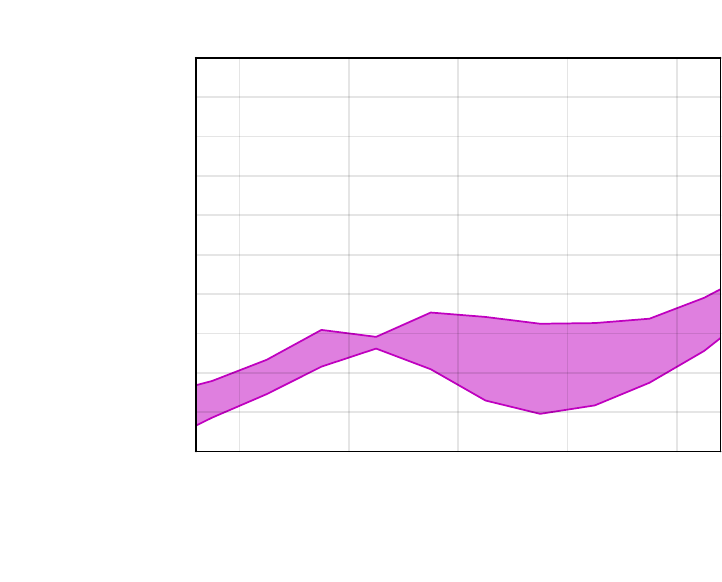}}\begin{picture}( 208.557970, 167.969210)\end{picture}\kern -208.557970pt\definecolor{ASYcolor}{gray}{0.000000}\color{ASYcolor}
\fontsize{12.045000}{14.454000}\selectfont
\usefont{\ASYencoding}{\ASYfamily}{\ASYseries}{\ASYshape}\ASYalignT(-107.739126,28.135020)(-0.500000,-1.000000){1.000000 0.000000 0.000000 1.000000}{\vphantom{$10^4$}$1$}\definecolor{ASYcolor}{gray}{0.000000}\color{ASYcolor}
\fontsize{12.045000}{14.454000}\selectfont
\ASYalignT(-44.510780,28.135020)(-0.500000,-1.000000){1.000000 0.000000 0.000000 1.000000}{\vphantom{$10^4$}$2$}\definecolor{ASYcolor}{gray}{0.000000}\color{ASYcolor}
\fontsize{12.045000}{14.454000}\selectfont
\ASYalignT(-76.124953,13.505625)(-0.500000,-0.750000){1.000000 0.000000 0.000000 1.000000}{$p_T$ (\si{GeV})}\definecolor{ASYcolor}{gray}{0.000000}\color{ASYcolor}
\fontsize{12.045000}{14.454000}\selectfont
\ASYalignT(-161.303020,37.439071)(-1.000000,-0.500000){1.000000 0.000000 0.000000 1.000000}{\vphantom{$10^4$}$0$}\definecolor{ASYcolor}{gray}{0.000000}\color{ASYcolor}
\fontsize{12.045000}{14.454000}\selectfont
\ASYalignT(-161.303020,60.201276)(-1.000000,-0.500000){1.000000 0.000000 0.000000 1.000000}{\vphantom{$10^4$}$0.002$}\definecolor{ASYcolor}{gray}{0.000000}\color{ASYcolor}
\fontsize{12.045000}{14.454000}\selectfont
\ASYalignT(-161.303020,82.963481)(-1.000000,-0.500000){1.000000 0.000000 0.000000 1.000000}{\vphantom{$10^4$}$0.004$}\definecolor{ASYcolor}{gray}{0.000000}\color{ASYcolor}
\fontsize{12.045000}{14.454000}\selectfont
\ASYalignT(-161.303020,105.725685)(-1.000000,-0.500000){1.000000 0.000000 0.000000 1.000000}{\vphantom{$10^4$}$0.006$}\definecolor{ASYcolor}{gray}{0.000000}\color{ASYcolor}
\fontsize{12.045000}{14.454000}\selectfont
\ASYalignT(-161.303020,128.487890)(-1.000000,-0.500000){1.000000 0.000000 0.000000 1.000000}{\vphantom{$10^4$}$0.008$}\definecolor{ASYcolor}{gray}{0.000000}\color{ASYcolor}
\fontsize{12.045000}{14.454000}\selectfont
\ASYalignT(-161.303020,151.250095)(-1.000000,-0.500000){1.000000 0.000000 0.000000 1.000000}{\vphantom{$10^4$}$0.01$}\definecolor{ASYcolor}{gray}{0.000000}\color{ASYcolor}
\fontsize{12.045000}{14.454000}\selectfont
\ASYalignT(-193.185755,94.344583)(-0.500000,0.189099){0.000000 1.000000 -1.000000 0.000000}{$v_3^\text{hfe}$}\definecolor{ASYcolor}{gray}{0.000000}\color{ASYcolor}
\fontsize{12.045000}{14.454000}\selectfont
\ASYalign(-76.124953,154.863595)(-0.500000,0.068966){\large \textsc{Rhic} $0$--$60$\%}     \caption{Calculation of electron $v_3$ for \textsc{Rhic} energy.}
    \label{fig:v3rhic}
  \end{minipage}
\end{figure}

Also, in Figures~\ref{fig:v2rhic} and \ref{fig:v3rhic} flow calculations for
the \textsc{Rhic} energy. The results for $v_2$ are below the data which
might be due to the energy loss model which we want to improve in order to
obtain better results, also the $p_T$ range of the data is fairly limited in
comparison with the available range for $R_\text{AA}$ spectra.

\section{Conclusions}

We implemented a 2D+1 Langrangian ideal hydrodynamic code on an
event-by-event basis and evolved heavy quarks inside the expanding medium in
order to obtain electron $p_T$ and $\varphi$ spectra so we could calculate
the nuclear modification factor $R_\text{AA}$ and the Fourier harmonic
coefficients. Our framework is intentionally very modular so one could use it
to test different models for energy loss, hydrodynamics or fragmentation,
allowing for a comprehensive study of different aspects of the \textsc{qgp}
phenomenology. We obtained $R_\text{AA}$ spectra for both \textsc{Rhic} and
\textsc{Lhc} energies.

\bibliography{library.bib}
\end{document}